\documentclass[aps,prb,twocolumn,superscriptaddress]{revtex4-2}

\usepackage{graphicx}
\usepackage{dcolumn}
\usepackage{bm}
\usepackage{hyperref}
\usepackage[moderate]{savetrees}
\begin{document}

\preprint{APS/123-QED}

\title{Observation and control of collective spin-wave mode-hybridisation in chevron arrays and square, staircase and brickwork artificial spin ices}

\author{T. Dion}
\email[]{troy.dion@phys.kyushu-u.ac.jp}
\affiliation{Solid State Physics Lab., Kyushu University, 744 Motooka, Nishi-ku, Fukuoka, 819-0395, Japan
}
\affiliation{London Centre for Nanotechnology, University College London, London WC1H 0AH, United Kingdom
}

\author{J. C. Gartside}
\author{A. Vanstone}
\author{K. D. Stenning}
\affiliation{%
Blackett Laboratory, Imperial College London, London SW7 2AZ, United Kingdom
}

\author{D. M. Arroo}
\affiliation{London Centre for Nanotechnology, Imperial College London, London SW7 2AZ, UK}
\affiliation{Department of Materials, Imperial College London, London SW7 2AZ, UK
}
\author{H. Kurebayashi}
\affiliation{London Centre for Nanotechnology, University College London, London WC1H 0AH, United Kingdom
}

\author{W. R. Branford}
\affiliation{Blackett Laboratory, Imperial College London, London SW7 2AZ, United Kingdom
}
\affiliation{London Centre for Nanotechnology, Imperial College London, London SW7 2AZ, UK}

\date{\today}

\begin{abstract}
Dipolar magnon-magnon coupling has long been predicted in nano-patterned artificial spin systems. However, observation of such phenomena and related collective spin-wave signatures have until recently proved elusive or limited to low-power edge-modes which are difficult to measure experimentally. Here we describe the requisite conditions for dipolar mode-hybridisation, how it may be controlled, why it was not observed earlier and how strong coupling may occur between nanomagnet bulk-modes. We experimentally investigate four nano-patterned artificial spin system geometries: `chevron' arrays, `square', `staircase' and `brickwork' artificial spin ices. We observe significant dynamic dipolar-coupling in all systems with relative coupling strengths and avoided-crossing gaps supported by micromagnetic-simulation results. We demonstrate reconfigurable mode-hybridisation regimes in each system via microstate control, and in doing so elucidate the underlying dynamics governing dynamic dipolar-coupling with implications across reconfigurable magnonics. We demonstrate that confinement of the bulk-modes via edge effects play a critical role in dipolar hybridised-modes, and treating nanoislands as a coherently precessing macro-spins or standing spin-waves are insufficient to capture experimentally-observed coupling phenomena. Finally, we present a parameter-space search detailing how coupling strength may be tuned via nanofabrication-dimensions and material properties.

\end{abstract}

\maketitle

\section*{Introduction}
Artificial spin ices (ASI) are arrays of magnetically frustrated nanoislands with vast low-energy state degeneracy \cite{wang2006artificial,Nisoli2013,skjaervo2020advances,gartside2018realization}. Study of ASI and related systems has expanded beyond modelling thermodynamic systems to leveraging them as host platforms for diverse applications including reconfigurable magnonics \cite{Kaffash2021,barman2020magnetization,wang2006artificial,lendinez2019magnetization,gliga2020dynamics,skjaervo2020advances,Talapatra2020,gartside2020current}, neuromorphic \cite{grollier2020neuromorphic,hon2021numerical} and reservoir computing \cite{nomura2021reservoir,jensen2020reservoir,gartside2021reconfigurable}. Bypassing the need for parity of interactions, differential fabrication \cite{dion2019tunable,Gartside2021,Vanstone2021,Lendinez2021} offers enhanced tunability of the dynamic magnon response and increased microstate access flexibility. 
Reconfigurable magnonic crystals (RMC) \cite{grundler2015reconfigurable,topp2010making,krawczyk2014review,haldar2016reconfigurable,barman2020magnetization,roadmap,Topp2011,Wang2009,Al-Wahsh2011,Ma2012,haldar2016reconfigurable,Gubbiotti2005,Gubbiotti2007,Wang2009,topp2010making,arroo2019sculpting} are highly-attractive due to hosting many distinct spin-wave spectra, with promising information processing applications \cite{kruglyak2010magnonics,lenk2011building}. Magnonic crystals can express spin-wave band gaps, band-pass filtering, and waveguide bending\cite{Al-Wahsh2011,roadmap}. Diverse functionality within the same RMC allows a plethora of different computational tasks and offers a potential solution to high power consumption and waste heat \cite{chumak2015magnon} associated with traditional CMOS electronics. An attractive RMC avenue is engineering dipolar magnon-magnon coupling between nanomagnets. Typically, coupled magnetisation dynamics is achieved via short-range exchange interaction \cite{shiota2020tunable,sud2020tunable}, placing tight-constraints on experimental system architecture. The dipolar-interaction responsible for coupling in nano-patterned RMC offers relative freedom and reconfigurability of mode-hybridisation phenomena \cite{gartside2021reconfigurable}. There are many demonstrations of RMCs using 1D arrays \cite{Wang2009,Al-Wahsh2011,Ma2012,haldar2016reconfigurable,Gubbiotti2005,Gubbiotti2007,Wang2009,topp2010making,Topp2011} which while impressive suffer from limited number of states versus 2D arrays. ASIs are attractive to magnonic computing since they can be leveraged more flexibly and exhibit richer spin-wave spectra \cite{Kaffash2021,arroo2019sculpting,dion2019tunable,gartside2021reconfigurable}. Rapid readout techniques for microstates have been developed for ASI making it a promising RMC candidate \cite{Gliga2013,Vanstone2021}.

Previously dipole-dipole coupling and collective spin-wave behaviour in ASI proved elusive and avoided-crossings had not been observed. Interactions were considered too weak to resolve in ASI bulk-modes (BM) \cite{Li2017} or limited to low-power edge-modes (EM) \cite{Heyderman,Kostylev2007} which are challenging to detect experimentally due to smaller magnetic volume and imperfect nano-patterned edges although can be improved using ion-beam milling \cite{Urbanek2010}. Simulation of coupled nanomagnets in ASI where inter-island coupling is mediated by spin-wave channels in an exchange-biased underlayer has been demonstrated \cite{Iacocca2020}. Here we show inter-island dipolar-coupling is sufficient for opening spin-wave band gaps using micromagnetic-simulation (MuMax3 \cite{vansteenkiste2014design}) and experimental ferromagnetic resonance (FMR).
 
We previously investigated width-modified bi-component square ASI, alternating rows of thin and wide nanoislands along each sublattice, termed `staircase' ASI. This provides access to `type-3' states consisting of `3-in, 1-out' vertex-configuration whose spin-wave signature had yet been measured. Applying field 45$^\circ$ to sublattice axes we observed an avoided-crossing due to anti-parallel magnetisation. This geometry with perpendicular state-preparation and measurement field-directions is atypical, and its use and efficacy in exploring mode-hybridisation is further investigated here. We show collective spin-wave modes are not limited to geometrically-modified or 1D arrays and present a detailed study elucidating contributing factors to hybridised spin-wave modes in strongly-interacting nanomagnetic arrays. We investigate `diagonal' and `chevron' two nanoisland arrays, `square', `staircase' and `brickwork' ASI. A systematic parameter search is performed, including nanoisland-dimensions, vertex gap, array geometry and saturation magnetisation for `square' ASI. The results shed light on dipolar magnon-magnon coupling and form a set of design-rules for tailoring and controlling dipolar hybridisation phenomena in artificial spin-system meta-materials \cite{Talapatra2020}. Sample fabrication, experimental, fitting and simulation methods are all found in supplementary information.

\section*{Results and Discussion}
\subsection*{Acoustic and optical spin-waves}

\begin{figure}[thbp] 
\centering
\includegraphics[width=0.45\textwidth]{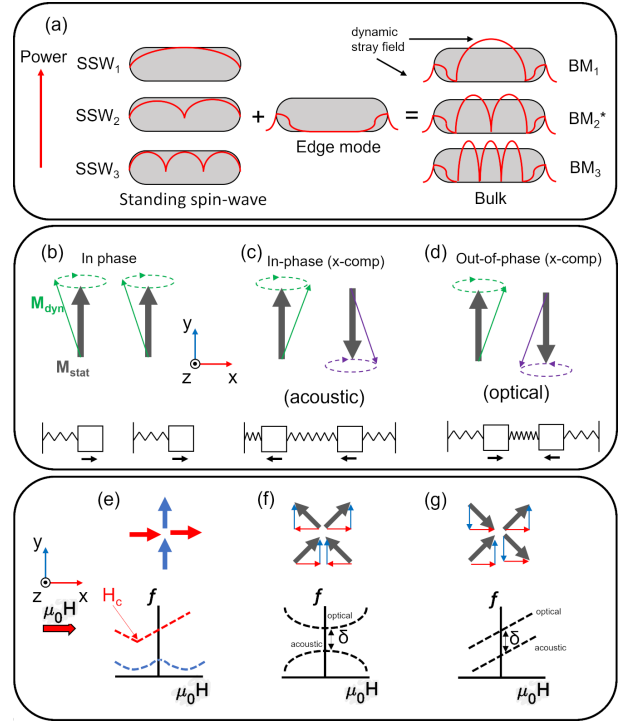}
\caption{(a) Power distributions of single nanoisland-modes; standing spin-wave (SSW) modes, edge-modes (EM) and bulk-modes (BM). BM observed experimentally and  simulation are a combination of EM and SSW. Stray-field can emanate from long edges and short edges. *not excited in uniform field.
(b) Two parallel-magnetised nanoislands precess coherently, analogous to two uncoupled masses (shown underneath). 
(c) Acoustic-mode where oppositely magnetised nanoislands' $x$-components of dynamic magnetisation precess in-phase, analogous to two spring-coupled masses moving coherently. 
(d) Optical-mode where oppositely magnetised nanoislands' $x$-components move out-of-phase, analogous to two spring-coupled masses moving in opposite directions. 
(e) `Type-2' ASI microstate with field applied parallel to sublattice.
(f) `Type-2' ASI magnetised perpendicular to the applied field direction. Hybridisation occurs only near avoided-crossing.
(g) `Type-2' ASI microstate with parallel preparation and measurement field.}
\label{Fig1} 
\end{figure}

Effective inter-island coupling requires significant dynamic stray-field. Figure \ref{Fig1}(a) illustrates different nanoisland spin-wave modes; standing spin-wave modes (SSW) which exhibit insignificant stray-field, EM predicted to exhibit mode-hybridisation due to stray-field and BM which exhibit an uncharacteristic combination of EM and SSW with stray-field emanating from both short and long edges of the nanoisland allowing opportunity for effective inter-island coupling.

When moments are aligned parallel ($\uparrow \uparrow$), no mode splitting occurs and both moments precess in-phase (Fig. \ref{Fig1}(b)). When aligned anti-parallel ($\uparrow \downarrow$), mode-hybridisation occurs and when the energies of BM in separate nanoislands are brought close together an avoided-crossing is observed. We know from studies on synthetic antiferromagnets (syAFM) \cite{sud2020tunable,shiota2020tunable,kalinikos1986theory,liensberger2019exchange} and bistable 1D nanoisland arrays \cite{topp2010making,haldar2016reconfigurable,Topp2011} that hybridised-modes are distinguished by in-plane dynamic magnetisation moving in-phase or out-of-phase termed acoustic and optical respectively as illustrated in Figs. \ref{Fig1}(c-d). The out-of-plane dynamic magnetisation has the opposite phase relationships, ie. acoustic moves out-of-phase and the optical moves in-phase due to opposite precession chirality. The frequency gap, $\delta$, is caused by Brillouin zone folding \cite{topp2010making,Topp2011}, typically occurring near remanance but can be shifted in field by width modification (`staircase') or symmetry breaking  (`brickwork').

No strong inter-island coupling exists if external fields are applied along sublattice directions \cite{Li2017}. Fig. \ref{Fig1}(e) illustrates the coercive field of the parallel-to-field nanoislands (red) are much lower than the perpendicular-to-field nanoislands (blue) and therefore the avoided-crossings cannot be observed in principle. Figures \ref{Fig1}(f-g) show applying the field diagonally brings modes close together and each nanoisland magnetisation can be broken down into two $x,y$-components. For hybridisation between neighbouring single-nanoisland modes to occur a pair of nanoislands must have $\uparrow \downarrow$ configuration; in Fig. \ref{Fig1}(g) $x$-component  $\rightleftharpoons$ (red) or Fig. \ref{Fig1}(f) the $y$-component $\uparrow \downarrow$ (blue). It follows that avoided-crossings should be observable in a field-saturated `type-2' state if measurement and preparation fields are perpendicular. Typically arrays are saturated along a given axis, then spectra measured while sweeping field along the same axis, as illustrate in Fig. \ref{Fig1}(e) \cite{Li2017,Lendinez2021,Li2016,jungfleisch2016dynamic}, partially explaining why avoided-crossings had not been observed before. The microstate in Fig. \ref{Fig1}(g) has modes with same field-gradient sign but should still exhibit a gap but since $\delta$ is typically on the order of hundreds of MHz it is likely obscured by experimental linewidth.

\subsection*{Mode-hybridisation in two-nanoisland arrays}

\begin{figure*}[thbp] 
\centering
\includegraphics[width=0.8\textwidth]{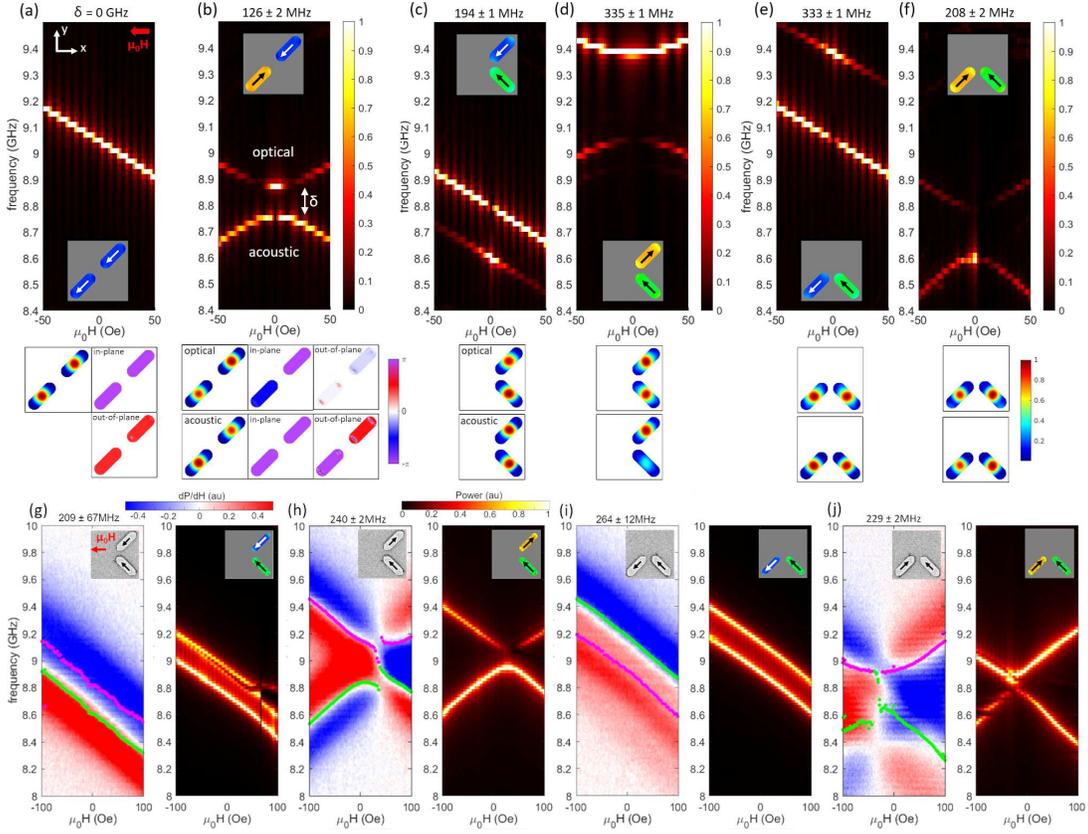}
\caption{
(a-f) Simulated spectra for `diagonal' (a,b) and `chevron' (c-f) two-nanoisland configurations. Microstates are shown inset for each plot, corresponding power and phase maps shown below. Applied field direction is along x-axis and swept from positive to negative as indicated by red arrow in (a). 
(g-j) Experimental (blue/white/red) and corresponding simulation (black/red/white) for each microstate/field configuration. Two peaks are not immediately resolvable in (g,i) but broader tails on high-frequency (g) and low-frequency sides (i) are evidence of a second lower-power peak, consistent with simulations (c) and (e). Lorentzian fits represented by green and pink dots (g-j). 
}
\label{Fig2} 
\end{figure*}

Three distinct two-nanoisland systems are shown in Fig. \ref{Fig2} with nanoisland-dimensions 220 by 80 by 20 nm and lattice parameter, $\Lambda$ = 300 nm. Spatial Fourier transforms applied to magnetisation time-series to generate spin-wave spectra, power and phase maps. The `diagonal' two-nanoisland system (Fig. \ref{Fig2}(a,b)) demonstrates clear distinction between the $\uparrow \uparrow$ (Fig. \ref{Fig2}a) and $\uparrow \downarrow$ (b) spectra. Figure \ref{Fig2}(a) exhibits a single mode, increasing in frequency as field is swept positive to negative along the x-axis. The corresponding power plot underneath shows equal power in both nanoislands and phase plots show both in-plane (top) and out-of-plane (bottom) magnetisation precessing in-phase. Figure \ref{Fig2}(b) shows the $\uparrow \downarrow$ case exhibiting acoustic and optical-modes with an avoided-crossing of $\delta$ = 126 $\pm$ 2 MHz at zero-field, relatively small due to the centre-to-centre distance being $\Lambda$ compared with `chevron' arrays with $\Lambda/\sqrt{2}$. Power maps appear similar, but phase maps reveal expected optical and acoustic-mode phase relationships. Acoustic-mode has in-plane components of magnetisation moving in-phase and optical-mode has in-plane components moving out-of-phase. 

Next we examine simulated spin-wave spectra of four possible microstates (Figs. \ref{Fig2}(c-f)) in a chevron' geometry. In Fig. \ref{Fig2}(c) the $x$-component of the magnetisation ($m_x$) is collinear with the swept magnetic field direction and $\delta$ is constant for all fields. Along the $y$-component of magnetisation ($m_y$) one nanoisland points up (+$m_y$) and the other down (-$m_y$), satisfying optical and acoustic-mode generation conditions. The two modes exhibit $\delta$ = 194 $\pm$ 1 MHz, higher than Fig. \ref{Fig2}(b) since the inter-island distance is $\Lambda/\sqrt{2}$. A large $\delta$ of 335 $\pm$ 1 MHz is observed in Fig. \ref{Fig2}(d), showing strong inter-island mode coupling. 

$\delta$ depends not only on microstate but also the local field which is a function of magnetisation alignment favourability. Figures \ref{Fig2}(d,e) are favourably aligned showing higher overall frequency and $\delta$ compared to Figs. \ref{Fig2}(c,f) which are unfavourably configured. Since the frequency increases with effective field, $\mathbf{H}_{\mathrm{eff}}$, which includes a dipolar-field term $\mathbf{H}_{\mathrm{dip}}$, the cancellation of the dipolar-fields when two moments are both pointing into the vertex lowers the resonant frequency. In Figs. \ref{Fig2}(c,e) the upper mode has higher or lower power respectively allowing experimental detection via asymmetry in the spin-wave signature. 

Figures \ref{Fig2}(g-j) are experimental (red/blue) and simulated (black/red/white) results for the fabricated `chevron' sample with dimensions 540 by 140 by 25 nm and $\Lambda$ = 800 nm. The experimental differential FMR heatmaps $(\frac{\partial P}{\partial H})$ are consistent with simulation. Larger $\delta$ is observed for microstates with favourable alignment in Figs. \ref{Fig2}(h,i) with $\delta$ = 240 $\pm$ 2 MHz and 264 $\pm$ 12 MHz respectively. Unfavourable microstates in Figs. \ref{Fig2}(g,j) have a smaller $\delta$ of 209 $\pm$ 67 MHz and  229 $\pm$ 2 MHz respectively demonstrating tunability via microstate. The relative optical and acoustic-mode power depends on the configuration as discussed above. The asymmetry of red and blue shading and Lorenztian fitting (green and pink dots) in Figs. \ref{Fig2}(g,i) reveal two modes.

For conventional symmetric ASI in `type-2' states mode shifting due to local field distributions is insignificant. For symmetry broken `type-3' ASI states, `brickwork', or width-modified ASI the vertex driven local fields can be leveraged to tune $\delta$ via field or microstate control. 

\subsection*{Mode-hybridisation in ASI}

\begin{figure*}[thbp]   
\centering
\includegraphics[width=0.95\textwidth]{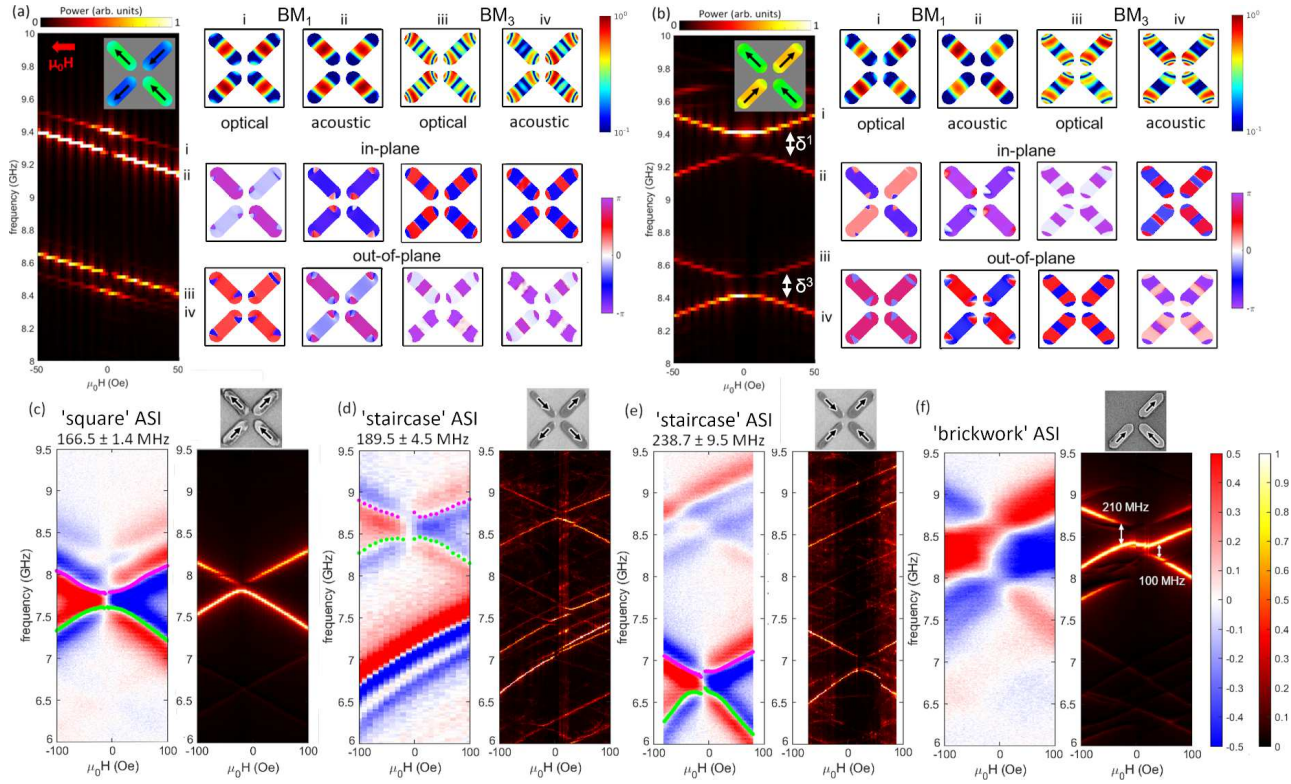}
\caption{(a) Spectra for the `type-2` microstate shown inset saturated along the swept field direction (indicated by red arrow). Modes i-iv are indicated on the right side of the spectra. Modes i and ii are similar to those in Fig. 1 with power concentrated in nanoisland centre. Modes $\textrm{BM}_{3}^{opt}$ and $\textrm{BM}_{3}^{aco}$ have shorter wavelength and backward-volume magnetostatic spin wave character (BVMSW). (b) Spectra for `type-2' microstate saturated perpendicular to the swept-field direction. (c-f) Experimental and simulation results for (c) `square' ASI, (d-e) `staircase' ASI and (f) `brickwork' ASI. Black arrows illustrate microstate}
\label{Fig3} \vspace{-1em}
\end{figure*}

Figure \ref{Fig3} shows simulated spectra for square ASI microstates saturated with the measurement field parallel (a) and perpendicular (b) to the preparation field. In simulations coupling for the higher order BM is observed and we define the two gaps as $\delta^1 = f(\textrm{BM}_{1}^{opt})- f(\textrm{BM}_{1}^{aco})$ = 146 $\pm$ 2 MHz and $\delta^3 = f(\textrm{BM}_{3}^{opt})- f(\textrm{BM}_{3}^{aco})$ = 127 $\pm$ 2 MHz. $\textrm{BM}_{3}$ are present in experiment but too faint to make quantitative assessment (see supplementary material).

Modes i and ii in Figs. \ref{Fig3}(a,b) follow similar phase relationship as the two-island case. Remembering that the $\textrm{BM}_{1}^{opt}$  has in-plane magnetisation of coupled-nanoislands moving out-of-phase with each other. For $\textrm{BM}_{1}^{aco}$ the phase of the precession of coupled-nanoislands move in-phase. $\textrm{BM}_{3}$ are best described by backward volume magnetostatic spin waves (BVMSW) \cite{kalinikos1986theory} where the wavevector and magnetisation are both defined parallel to the nanoisland long-axis. $\textrm{BM}_{3}^{opt}$ and $\textrm{BM}_{3}^{aco}$ have lower frequency than the $\textrm{BM}_{1}$, as expected for BVMSW \cite{Chumak2008}. In Fig. \ref{Fig3}(b) the lower power of $\textrm{BM}_{1}^{aco}$ is due to homogeneous excitation field inefficiently driving a mode where out-of-phase precession is the resonant condition. 

Experimentally-measured and simulated FMR spectra for four ASI cases exhibiting collective spin-wave signatures are compared; square ASI in a perpendicular `type-2' state, `staircase' ASI \cite{gartside2021reconfigurable} in `type-3' states; wide-nanoisland and thin-nanoisland majority magnetisation and perpendicularly-saturated `brickwork' ASI (Fig. \ref{Fig3}(f)).


Figure \ref{Fig3}(c) shows spectra for symmetric `square' ASI with dimensions 460 by 150 by 25 nm and $\Lambda$ = 600 nm. $\delta$ = 166.5 $\pm$ 1.4 MHz, demonstrating avoided-crossings are experimentally resolved in ASI without differential fabrication.


Figures \ref{Fig3}(d-e) show spectra for a `staircase` ASI with dimensions 600 by 200 (wide) / 130 (thin) by 20 nm and $\Lambda$ = 800 nm. Figure \ref{Fig3}(d), shows  $\delta^1$ = 189.5 $\pm$ 4.45 MHz, and  \ref{Fig3}(e), $\delta^1$ = 238.7 $\pm$ 9.5 MHz. Preparing the wide-nanoisland majority `type-3' state locates the avoided-crossing in the high-frequency, hybridised thin-nanoisland modes. The thin-nanoisland majority `type-3' exhibits avoided-crossing in the low-frequency, wide-nanoisland hybridised modes. This demonstrates tunability of $\delta^1$ via microstate switching for RMC applications.


Figure \ref{Fig3}(f) shows `brickwork' ASI with dimensions 570 by 170 by 25 nm and $\Lambda$ = 800 nm achieved via single nanoisland removal from the unit cell. The simulated avoided-crossing gaps were $\delta^1$ = 240 $\pm$ 25 MHz  in the negative field region (right coupled nanoislands) and $\delta^1$ = 100 $\pm$ 25 MHz in the positive field region (bottom coupled nanoislands). Unfortunately accurate experimental peaks were not extracted due to larger linewidth, however, the presence of avoided-crossings is plausible based on the similarity to the simulated spectra. Observing two different $\delta^1$ is consistent with the `chevron' findings and allows tunability via field without microstate change. Lorentzian fits to experimental data are available in supplementary material.

\subsection*{Mode profiles}

\begin{figure*}[htbp]   
\centering
\includegraphics[width=0.95\textwidth]{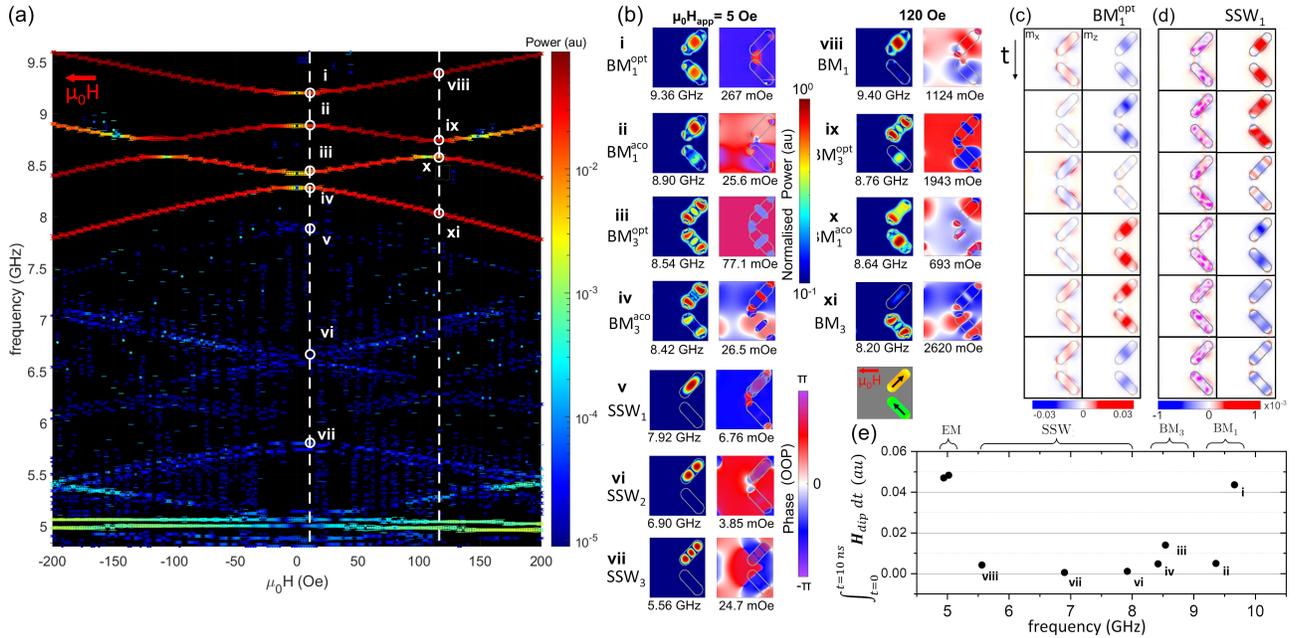}
\caption{(a) Peak extraction of $\pm$20 mT high frequency-resolution sweep for same chevron state in Fig. \ref{Fig2}(d). Avoided-crossings are observed at higher fields between modes $\textrm{BM}_{1}$ (ix) and $\textrm{BM}_{3}$ (x). (b) Power and out-of-plane phase maps for each of the labelled modes at 5 Oe and 120 Oe also indicated by white dotted lines in (a). Analysis is applied to demagnetising field to visualise dynamic stray-field. Stray-field value is displayed under phase maps. Time-domain simulations for (c) $\textrm{BM}_{1}^{opt}$ and (d) SSW$_{1}$ mode where one precession cycle is shown. Precession for SSW$_{1}$  is purely out-of-plane. $\textrm{BM}_{1}^{opt}$ shows more coherent in-plane mode structure and is a combination of EM and SSW. (e) Stray-field outside nanoisland is integrated over time and plotted for each mode.}
\label{Fig4} \vspace{-1em}
\end{figure*}

Spatial Fourier transforms are applied to the demagnetisation field to show dynamic stray-field profiles. Peak extractions are plotted in Fig. \ref{Fig4}(a) to better highlight weak peaks. $\textrm{BM}_{1}$ and $\textrm{BM}_{3}$ also exhibit an avoided-crossing at 120 Oe as indicated by ix and x. At this point $\textrm{BM}_{3}$ is optical and $\textrm{BM}_{1}$ is acoustic. BM revert to single nanoisland-mode behaviour at high fields.  Figure \ref{Fig4}(b) shows the corresponding power and phase of each mode labeled in Fig. \ref{Fig4}(a). $\mathrm{SSW}_{1-3}$ have significantly lower dynamic dipole-field confined to nanoisland volume agreeing with the theoretical assumption that no coupling for $\mathrm{SSW}_{1-3}$ is expected. Values shown underneath each phase plot in Fig. \ref{Fig4}(b) are dipole-field calculated outside nanoislands. Modes v and vi (non-coupled SSW) show significantly smaller values than i-iv (coupled BM). $\mathrm{BM}_{1}$ and $\mathrm{BM}_{3}$ power is clearly distributed across multiple nanoislands, contrasted with SSW localised to single nanoislands. 

The SSW picture applies well to low frequency modes that are not experimentally detected. Experimentally detected BM are best described by the combination of EM and SSW indicating geometry, particularly the nanoisland ends, plays a significant role. Additionally, there is a clear manifestation of stray-field emanating from long nanoisland edges. 

We excite modes sinusoidally to examine time-domain dynamics. Snapshots of demagnetising fields over a full precession cycle are shown in Figs. \ref{Fig4}(c,d) for the $\textrm{BM}_{1}^{opt}$ and SSW$_{1}$ modes respectively. All modes are available in supplementary video 1. $\textrm{BM}_{1}^{opt}$ shows coherent dynamics for the in-plane components whereas mode SSW$_{1}$ shows precession in $m_z$ only. Coherent in-plane precession for $\textrm{BM}_{1}^{opt}$ fosters inter-nanoisland coupling. The video shows the EM and SSW components of the BM have a transverse and longitudinal character respectively, seemingly arising due to the curved geometry at nanoisland ends.

Figure \ref{Fig4}(e) shows stray-field outside magnetic volume integrated over 10 ns. $\textrm{BM}_{1}^{opt}$ has comparable stray-field to the two EM further indicating edges are vital to BM coupling. Increasing the node number for SSW allows more stray-field to escape for potential coupling to occur, however, even in the simulations SSWs are far too weak to be experimentally detected.

\subsection*{Tailoring coupling via geometry}

\begin{figure*}[htbp]
\centering
\includegraphics[width=0.95\textwidth]{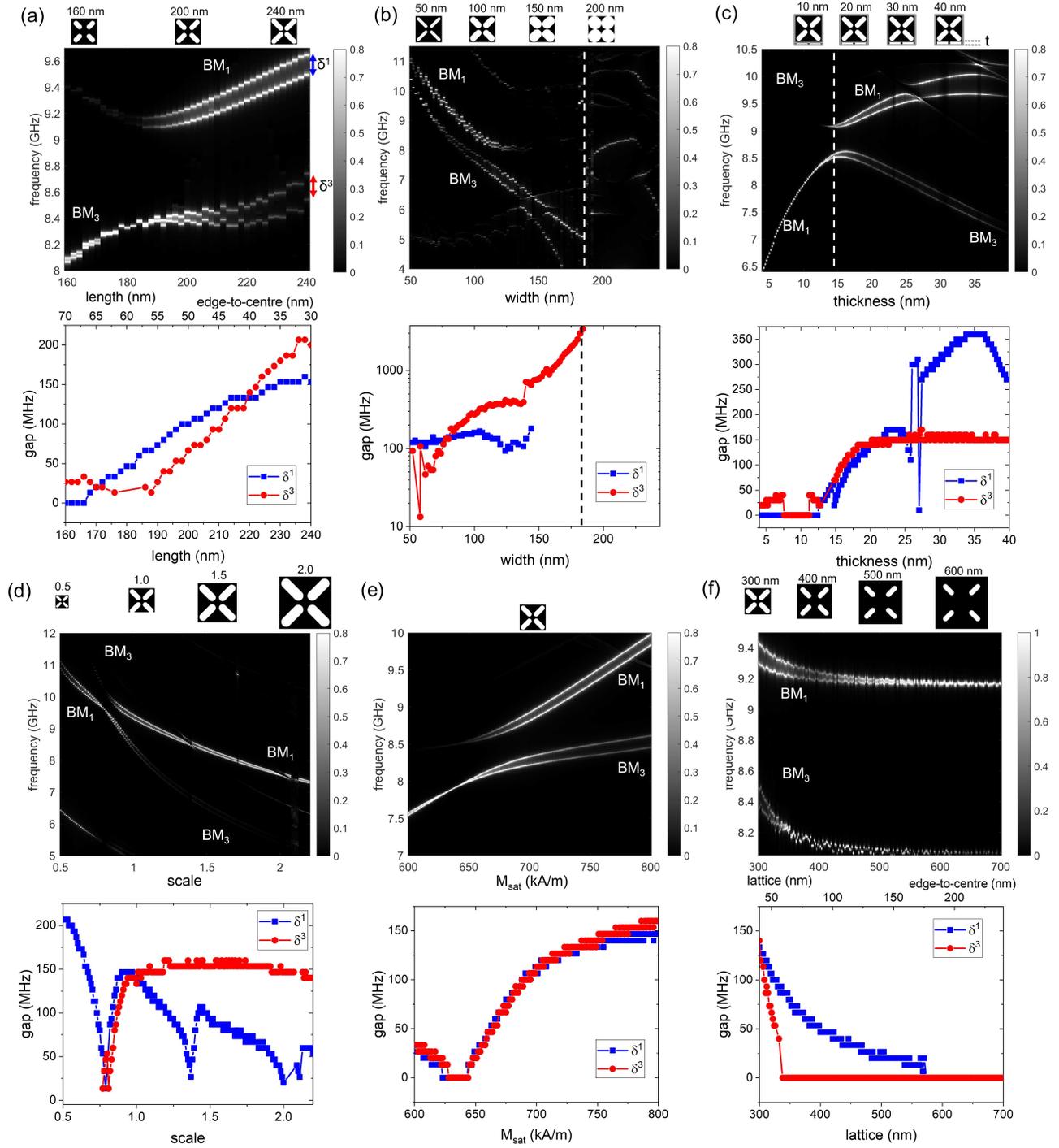}
\caption{Upper panels show the spectra and lower panels show the extracted $\delta$ for the $\delta^0 (f_i -f_{ii})$ and $\delta^1 (f_iii -f_{iv})$ modes as a function of (a) nanoisland length, (b) width, (c) thickness, (d), lateral scaling, (e) saturation magnetisation and (f) lattice parameter. Insets show geometry being changed and are all to scale with (e) showing geometry common to all panels. The dotted line in (c) indicates the point where the $\textrm{BM}_{1}^{opt}$ and $\textrm{BM}_{1}^{aco}$ modes switch position in frequency with the higher order $\textrm{BM}_{3}^{aco}$ and $\textrm{BM}_{3}^{opt}$ modes. This may be treated as another avoided-crossing. 
}
\label{Fig5} 
\end{figure*}

Tunability of frequency gaps between $\textrm{BM}_{1}$ ($\delta^1$) and  $\textrm{BM}_{3}$ ($\delta^3$) modes via parameter exploration is investigated using micromagnetic-simulation. The control case is a square ASI with 220 by 80 by 20 nm $\Lambda$ = 300 nm and $M_{sat}$ = 750 kA/m. 

Figure \ref{Fig5}(a) shows increasing nanoisland length 160-240 nm increases coupling strength 0-200 MHz. This is a function of nanoisland ends being in closer proximity and larger magnetic volume. Wave-vector may also play a role in determining how much stray-field escapes the nanoisland.  

Figure \ref{Fig5}(b) shows $\delta^1$ as a function of nanoisland width which remains relatively constant. $\delta^1$ disappears at 146 nm due to distortions arising from an ill-defined magnetisation vector (see supplementary). Wider nanoislands prevent the typical BM structure observed elsewhere, behaving more like a macro-spin with decreased stray-field. Interestingly $\delta^3$ depends strongly on width. The multi-nodal structure of the higher-order mode seems more robust and maintains its stray-field. The anti-node edge-proximity increases coupling significantly. The dotted line indicates the point at which neighbouring nanoislands become connected and different mode behaviour is observed. 

Figure \ref{Fig5}(c) shows how $\delta$ varies with thickness. A minimum thickness of $\sim$14 nm is required for measurable coupling, above which $\delta^1$ and $\delta^3$ increase up to constant value at around 20\% of nanoisland width, showing additional modes at large thicknesses. Intersection with a high-order EM interrupts $\delta^1$ for 25-27 nm and continues to increase thereafter up to 35 where additional modes interfere with the $\textrm{BM}_{1}$. There is an avoided-crossing at 15 nm, indicated by the dotted line. Figure \ref{Fig4} previously demonstrated a frequency-gap occurs between $\textrm{BM}_{3}$ and $\textrm{BM}_{1}$. For 14 nm and below the $\textrm{BM}_{3}$ have higher frequency than $\textrm{BM}_{1}$. Each of these avoided-crossings are explained by mode hybridisation confirmed by significant stray-field (see supplementary information). Below 14 nm $\textrm{BM}_{3}$ resemble magnetostatic surface spin-waves (MSSW) where they exhibit higher frequency than $\textrm{BM}_{1}$ \cite{Chumak2008}.

Figure \ref{Fig5}(d) shows $\delta$ varying with lateral scaling achieved by in-plane cell-size increase. $\delta^1$ decreases linearly with lateral scaling as volume increases as the square and dipole-dipole strength decreases as the cube resulting in a linear decrease overall. However, $\delta^1$ is significantly decreased when intersecting with other modes at 0.8 and 1.4 and 2.0. $\delta^3$ seems much less sensitive to the scaling parameter than $\delta^1$ above 1. As can be seen in Fig. \ref{Fig4}(b) maximal precession location of $\textrm{BM}_{3}$ is closer to nanosisland ends. The magnetic volume increases more than the coupling distance therefore exhibiting a much more shallow drop compared to $\delta^1$. 

Figure \ref{Fig5}(e) shows $\delta^1$ and $\delta^3$ both initially increase rapidly with $M_{sat}$ with diminishing returns. These findings imply that materials with higher saturation magnetisation like CoFeB could express significantly larger spin-wave band gaps. 

Finally Fig. \ref{Fig5}(f) shows how $\delta^1$ and $\delta^3$ decrease rapidly with increasing lattice parameter. This is strong evidence that the coupling is dipole mediated as it decreases with the cube and modes become decoupled when the inter-island distance is greater than about twice the nanoisland length.

\section*{Conclusion}

We investigate and explain the origin of avoided-crossings as hybridisation of single nanoisland-modes into acoustic and optical-modes which can be distinguished by precession phase relationships. Hyrbidisation between $\textrm{BM}_{1}$ and $\textrm{BM}_{3}$ also occurs at higher fields.

Avoided-crossings in artificial spin systems previously remained elusive due to typical experimental approaches employing preparation and measurement fields along the same axis. Rotating the applied field perpendicular to the preparation field or preparing broken-symmetry microstates (i.e. `type-3') generates a clear gap, $\delta$. Conversely, parallel preparation and detection field direction exhibit, $\delta$, often hidden by relatively large experimental line-widths. Theoretically it was assumed that insufficient stray-field between nanoislands prevents effective coupling. We show using simulation that this holds for SSW modes (Figs.\ref{Fig4}(b) v-vii). However, experimentally detected BM are a combination of SSW and EM, exhibiting significant stray-field shown in Fig. \ref{Fig4}(b) i-iv and avoided-crossings are confirmed via FMR. Nanoisland ends play a significant role in determining dipolar-coupling strength even between bulk-modes. 

The work presented here progresses the understanding of dynamic dipole-dipole coupling in nanomagnetic arrays, both its fundamental origin and how to harness and design it into systems. Much proposed magnonic computing is based on spin-wave interference effects in continuous magnetic media, here we show discrete nano-patterned islands can strongly interfere through collective-hybridisation - allowing interference effects with the reconfigurability and flexibility benefits of 2D-RMC. The nascent magnonic computing field requires heuristics of how to optimise systems for maximum interference, coupling and nonlinearity \cite{roadmap}. The design rules presented here provide means for this in nano-patterned 2D ASI-based RMC. There are multiple ways of tuning $\delta$ in the same structure via field application or microstate selection proving ASI and related structures to be promising candidates. We hope that unveiling the ability to generate and observe these avoided-crossings will encourage future studies into dipolar magnon-magnon coupling and collective mode-hybridisation in artificial spin systems and other architectures.


\subsection*{Acknowledgements}
TD supported by International Research Fellow of Japan Society for the Promotion of Science (Postdoctoral Fellowships for Research in Japan). WRB  supported by the Leverhulme Trust (RPG-2017-257). AV supported by the EPSRC Centre for Doctoral Training in Advanced Characterisation of Materials (Grant No. EP/L015277/1). TD and JCG conceived the work. JCG, KDS and AV fabricated the samples. AV and JCG performed CAD design of the structures. JCG, AV and KDS performed FMR measurements. JCG and AV performed analysis of FMR measurements. TD performed simulations and analysis. HK and DMA supported analysis of spin-wave spectra and interpretation. TD drafted the manuscript, with contributions from all authors in editing and revision stages. Simulations were performed on the Imperial College London Research Computing Service\cite{hpc}. The authors would like to thank Professor Lesley F. Cohen of Imperial College London for enlightening discussion and comments, and David Mack for excellent laboratory management. The authors declare no competing interests. Code and data-sets generated during and/or analysed during the current study are available from the corresponding author on reasonable request.

\bibliography{references}

\end{document}


\preprint{APS/123-QED}

\title{Supplementary Material -  Observation and control of collective spin-wave mode hybridisation in chevron arrays and square, staircase and brickwork artificial spin ices}

\author{T. Dion}
\email[]{troy.dion@phys.kyushu-u.ac.jp}
\affiliation{Solid State Physics Lab., Kyushu University, 744 Motooka, Nishi-ku, Fukuoka, 819-0395, Japan
}
\affiliation{London Centre for Nanotechnology, University College London, London WC1H 0AH, United Kingdom
}

\author{J. C. Gartside}
\author{A. Vanstone}
\author{K. D. Stenning}
\affiliation{%
Blackett Laboratory, Imperial College London, London SW7 2AZ, United Kingdom
}

\author{D. M. Arroo}
\affiliation{London Centre for Nanotechnology, Imperial College London, London SW7 2AZ, UK
}
\affiliation{Department of Materials, Imperial College London, London SW7 2AZ, UK
}
\author{H. Kurebayashi}
\affiliation{London Centre for Nanotechnology, University College London, London WC1H 0AH, United Kingdom
}

\author{W. R. Branford}
\affiliation{Blackett Laboratory, Imperial College London, London SW7 2AZ, United Kingdom
}
\affiliation{London Centre for Nanotechnology, University College London, London WC1H 0AH, United Kingdom
}

\date{\today}
             
 \maketitle
 
      \newcommand{\beginsupplement}{%
        \setcounter{table}{0}
        \renewcommand{\thetable}{S\arabic{table}}%
        \setcounter{figure}{0}
        \renewcommand{\thefigure}{S\arabic{figure}}%
     }
\beginsupplement 
 
 \section*{Methods}
\textbf{Frequency domain simulations:} Simulations are all performed in mumax3 \cite{vansteenkiste2014design} which solves the Landau-Lifshitz-Gilbert equation

\begin{widetext}
\begin{equation}
\label{LLG}
\frac{\partial \vec{M}(\vec{r},t)}{\partial t} = - \frac{\gamma}{1+\alpha^{2}} \vec{M}(\vec{r},t) \times \mu_0\vec{H}_{eff}(\vec{r},t) - \frac{\alpha \gamma}{M_{s}(1+\alpha^{2})} \vec{M}(\vec{r},t) \times (\vec{M}(\vec{r},t) \times \mu_0\vec{H_{eff}}(\vec{r},t))
\end{equation}
\end{widetext}

where $\vec{M}$ is the magnetisation,${M_s}$ is saturation magnetisation, $\gamma$ is the gyromagnetic constant, $\vec{\text{H}_{\text{eff}}}$ is the effective field and $\alpha$ is the Gilbert damping parameter.

The nanoislands are arranged on a 45$^{\circ}$ rotated square lattice where the number of islands can form a 2-island, 3-island (`brickwork') or 4-island (ASI) array allowing minimal unit cell for computation efficiency. Periodic boundary conditions simulate a quasi-infinite array. The magnetisation state is set manually and relaxed in an applied field to find the energy minima. A time-dependent sinc field pulse is applied in z-direction to mimic a broadband excitation. The simulation evolves in time according to the LLG equation and the (de)magnetisation is recorded at regular intervals. Dynamics  are obtained by subtracting ground state (t=0) from all subsequent time-steps. The data-set is windowed with a Hanning function (to reduce spectral leakage) before a fast Fourier transform (FFT) is applied along the time axis. Spatial power and phase plots are produced by integrating frequency bins in accordance with full width half maxima (FWHM) of peaks in the spectra. Each spatial power map colour-scale is normalised to the highest power pixel within the map. All simulations use the following parameters; magnetisation saturation $M_{sat}$ = 750kA/m, exchange stiffness $A_{ex}$ = 13 pJ, pulse amplitude = 50 Oe and time-step, $\Delta t$ = 1/(2*f). Variable simulation parameters are summarised in table \ref{paramtable}.

\begin{table*}[htbp]
\begin{center}
\begin{tabular}{||c c c c c c c c||} 
 \hline
Figure	&	Type	&		$c_xy$ (nm)	&	$c_z$ (nm)	&	lattice (nm)	&	dimensions (nm)	&	$f_{cut}$ (GHz)	&	$t_{tot}$ (ns)
 \\ [0.5ex] 
 \hline\hline
 2(a-f)	&	2-island	&		2.5	&	10	&	300	&	220$\times$80$\times$20	&	20	&	50
 \\ 
 \hline
 2(g-i)	&	2-island	&		5	&	20	&	800	&	540$\times$140$\times$25	&	10	&	50						 \\ 
 \hline	
3(a-b)	&	ASI	&		2.5	&	10	&	300	&	220$\times$80$\times$20	&	10	&	50							 \\ 
 \hline	
 3(c)	&	ASI	&		5	&	20	&	600	&	460$\times$150$\times$20	&	10	&	50						 \\ 
 \hline	
 3(d-e)	&	`staircase'		&	5	&	10	&	800	&	600$\times$(200/130)$\times$20*	&	20	&	100						 \\ 
 \hline		
3(f)	&	`brickwork'	&		5	&	10	&	800	&	570$\times$170$\times$25	&	10	&	100						 \\ 
 \hline		
5(a) 	&	ASI (length)	&		2	&	10	&	300	&	(160--240)$\times$120$\times$20**	&	10	&	150					 \\ 
 \hline	

5(b) 	&	ASI (width)	&		2	&	10	&	300	&	220$\times$(50--140)$\times$20**	&	15	&	150					 \\ 
 \hline			
5(c) 	&	ASI (thickness)	&		2	&	10	&	300	&	220$\times$80$\times$(4--40)**	&	10	&	150							 \\ 
 \hline	
5(d) 	&	ASI (scale)	&		2*(0.8--2)$\dagger$	&	10	&	300	&	(220$\times$80$\times$20)	&	15	&	150				 \\ 
 \hline		
 5(e) 	&	ASI ($\textrm{M}_{\textrm{sat}}$)	&		2	&	10	&	300	&	(220$\times$80$\times$20)	&	10	&	150				 \\ 
 \hline	
 5(f) &	ASI ($\Lambda$)	&			2	&	10	&	300 - 700	&	220$\times$80$\times$20	&	10	&	150							 \\ 
 \hline	

\end{tabular}
\caption{Variable parameters for simulations used throughout the paper. *Wide and narrow islands. ** Numbers in brackets shows minimum and maximum values. $\dagger$ Dimensions of Fig. 5(d) are scaled by increasing lateral cell size.}
\label{paramtable}
\end{center}
\end{table*}

\textbf{Time domain simulations:} To obtain the mode profiles in the time domain sinusoidal field is applied, ramped up to maximum amplitude using a tanh function, preventing excitation of high frequency modes. Data is saved every 2 ps for 10 ns. A region outside of the magnet is defined to quantify stray field generated at resonance. All components of this field are summed in quadrature and divided by the simulation volume.\\

\textbf{Sample fabrication:} Experimental samples were fabricated via E-beam liftoff lithography with PMMA resist and thermal evaporation of NiFe.\\

\textbf{Experimental measurement:} Experimental FMR measurements were performed using the commercial NanOsc cryoFMR probe in a Physical Properties Measurement System (PPMS) at room temperature. The samples were mounted `flip-chip' with the large ($\sim4 \times 4$ mm) array of ASI taped face down on the waveguide. Sample were orientated with applied field 45$^\circ$ to the islands axis and $\perp$ 'type 2' microstates were prepared prior to mounting in an external field. The applied field was modulated by a Helmholtz coil field (0.48 mT) at 490 Hz. The output from the waveguide is then rectified by an RF-diode and processed in a lock-in amplifier at the Helmholtz frequency. This produced the differential output shown in the normalised spectral heatmaps as a function of field and frequency. 
Experimental resonance frequencies were determined from peak fitted with derivative of a symmetric Lorentzian function using equation \ref{Lor}.
\begin{equation}
\label{Lor}
\frac{dP}{dH}(f) = A \cdot \frac{(\Delta f/2)(f-f_{res})}{((f-f_{res})^2+(\Delta f/2)^2)^2}
\end{equation}
where A is the differential amplitude, $\Delta f$ is the linewidth, and $f_{res}$ is the resonance frequency used for peak positions on the heatmaps.\\

\section*{Results} 

 \begin{figure*}
	\centering
		\includegraphics[width=0.6\textwidth]{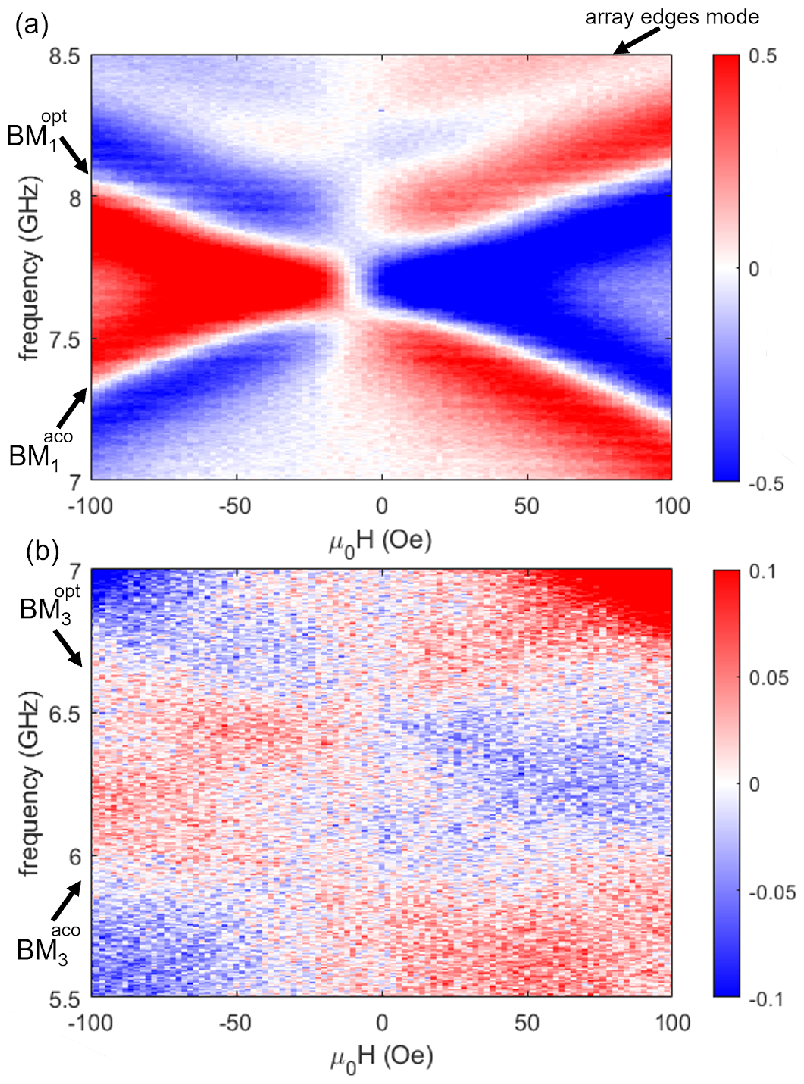}
	\caption{(a) Experimental spectral heatmap for 'square' ASI sample showing $\textrm{BM}_{1}^{opt}$ and $\textrm{BM}_{1}^{aco}$ modes. Due to narrowing of island width at array edges a faint mode at higher frequency is apparent. (b) Modes $\textrm{BM}_{3}^{opt}$ and $\textrm{BM}_{3}^{aco}$ are barely visible so not quantiative analysis is performed.}
	\label{S1}
\end{figure*}  

Figure \ref{S1} is a wider frequency window to show that the $\textrm{BM}_{3}$ modes are present in the sample but do not exhibit sufficient power to make a quantitative assessment. With more sensitive measurement techniques these modes should be detectable and this justifies our analysis of said modes in the simulations in the main manuscript of the paper. We also point out the array edge-modes. The width of nanoislands at the edges of the array have smaller width and therefore a mode with higher frequency is observed. 

\begin{figure*}
	\centering
		\includegraphics[width=1\textwidth]{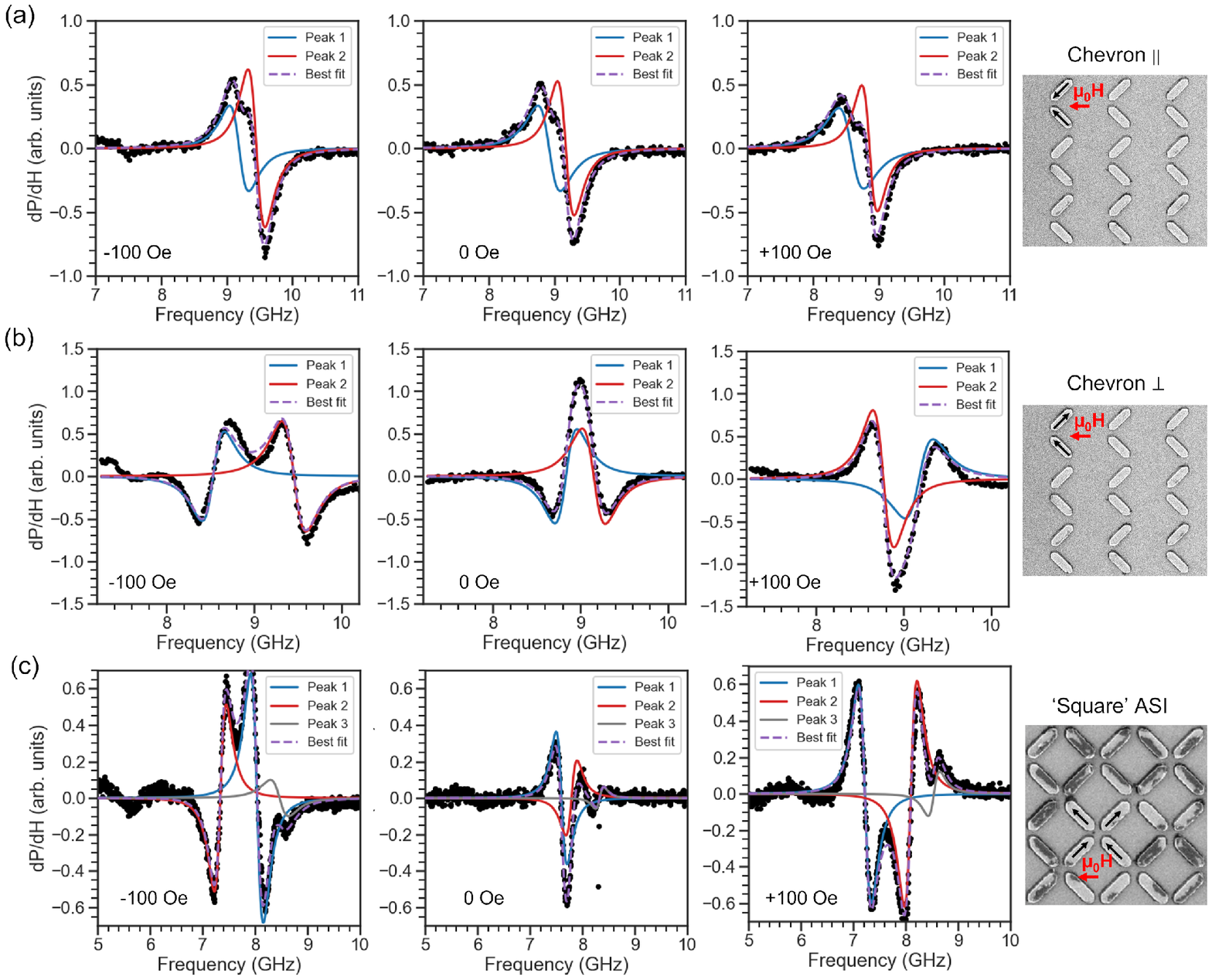}
	\caption{(a) Differential Lorentzian peak fits for the `chevron' sample in the parallel orientation (Fig. 2g). The component fitted peaks associated with the negatively magnetised islands acoustic mode (peak 1 – blue) and optical mode (peak 2 – red). (b) Differential Lorentzian peak fits `chevron' sample in perpendicular orientation (fig. 2h). The component fitted peaks associated with the negatively magnetised islands acoustic mode (peak 1 – blue) and optical mode (peak 2 – red). (c) Differential Lorentzian peak fits in fields for the `square' ASI sample (fig. 3c). The component fitted peaks associated with the negatively magnetised islands (peak 1 – blue), positively magnetised islands (peak 2 – red), and sample edge islands with a slightly smaller width (peak 3 – grey). The summation of the components is shown by the purple dashed line. Black arrows in SEM image show magnetisation state and red arrow shows the applied magnetic field direction.}
	\label{S2}
\end{figure*}

\begin{figure*}
	\centering
		\includegraphics[width=1\textwidth]{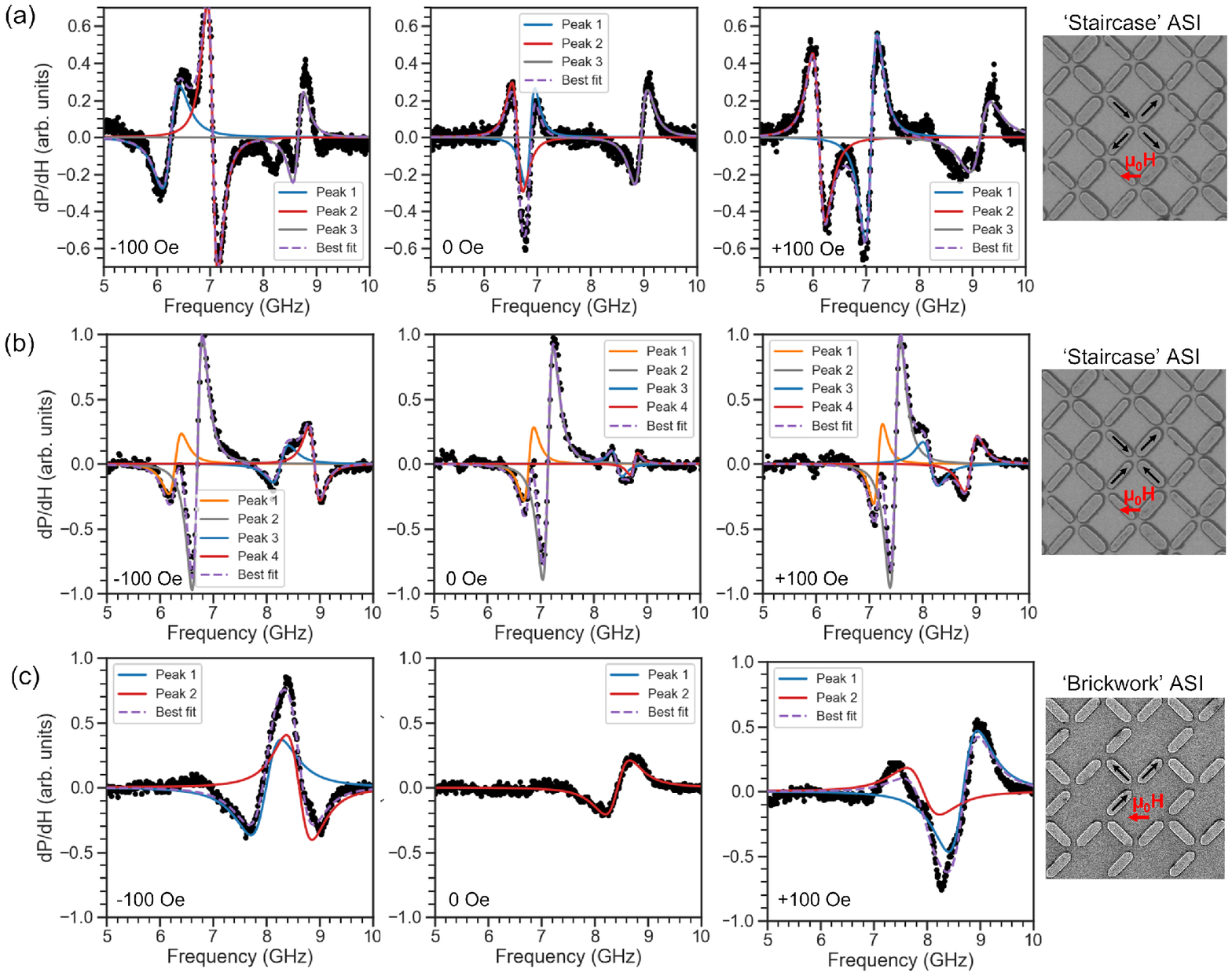}
	\caption{Differential Lorentzian peak fits in fields 'staircase' ASI sample prepared in wide majority state (fig. 3d). The component fitted peaks associated with the negatively magnetised bars (peak 1 – blue), positively magnetised bars (peak 2 – red), and negatively magnetised thin bars (peak 3 – grey). Differential Lorentzian peak fits for the 'staircase' ASI sample prepared in wide majority state (fig. 3d). The component fitted peaks associated with the negatively magnetised bars secondary bulk mode (peak 1 – orange), negatively magnetised bars bulk mode (peak 2 – grey), negatively magnetised thin bars (peak 3 – blue), and positively magnetised wide bars (peak 4 – red). Differential Lorentzian peak fits for the `brickwork' ASI sample (fig. 3f). The component fitted peaks associated with the negatively magnetised bars (peak 1 – blue) and positively magnetised bars (peak 2 – red). The summation of the components is shown by the purple dashed line. Black arrows in SEM image show magnetisation state and red arrow shows the applied magnetic field direction.}
	\label{S3}
\end{figure*}

Figures \ref{S2} and \ref{S3} are the Lorentzian peak fittings for each sample at -100, 0 and +100 Oe. The presence of two peaks in the 'chevron' sample in the parallel configuration demonstrates that the two peaks can be identified and the summation fits well to the experimental data. 

We also present power maps to complement parameter searches from Fig. 5 in the main manuscript.

\begin{figure*}
	\centering
		\includegraphics[width=0.6\textwidth]{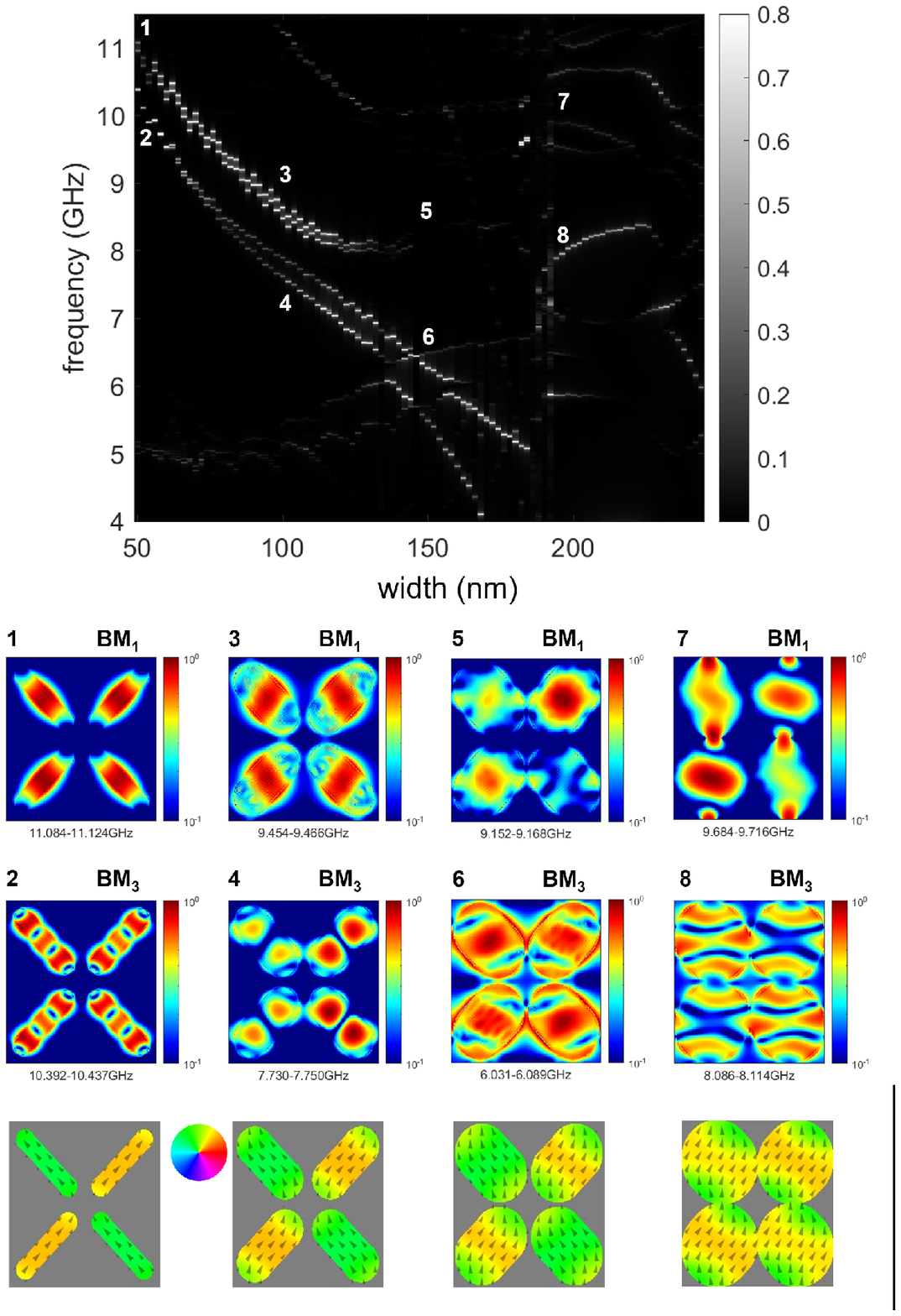}
	\caption{Power maps for Fig. 5b in the main manuscript in which the island width is investigated. Numbers correspond to modes in the heat map. Magnetisation texture is shown underneath colour corresponding to the colour wheel. Islands wider than 150 nm the $\textrm{BM}_{1}$ mode becomes distorted due to curved magnetisation texture. There is no clear magnetisation vector direction. Mode behaviour completely changes when neighbouring islands are proximate. $\textrm{BM}_{3}$ is more robust with increasing width. Increased node number seems to find a preferential direction despite curved magnetisation texture.}
	\label{S4}
\end{figure*}

Figure \ref{S4} shows how widening the islands causes distortions to the $\textrm{BM}_{1}$ profile but $\textrm{BM}_{3}$ remains robust up until the islands are no longer separated. There does not appear to be a clear direction preference for a single anti-node mode when the shape anisotropy is low but multi-nodal structures can form along the loose shape anisotropy that remains. 

\begin{figure*}
	\centering
		\includegraphics[width=0.8\textwidth]{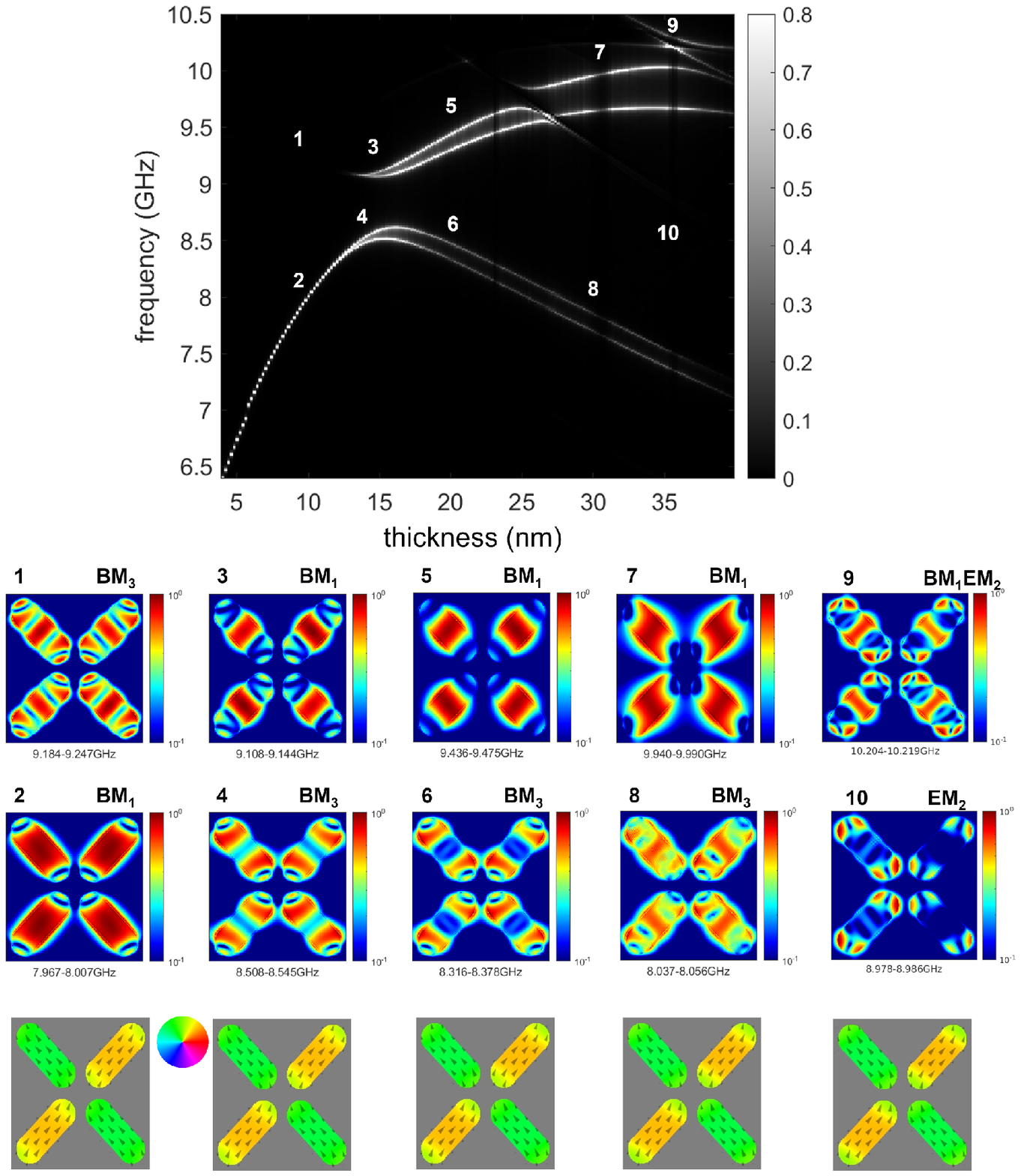}
	\caption{Power maps for Fig. 5c in the main manuscript in which the island thickness is investigated. Numbers correspond to modes in the heat map. Magnetisation texture is shown underneath colour corresponding to the colour wheel. Below 15 nm the $\textrm{BM}_{3}$ mode has higher frequency than $\textrm{BM}_{1}$ MSSW behaviour and BVMSW behaviour above. Larger thickness exhibits more magnetisation curling at the ends seemingly allowing modes to form along the out-of-plane axis of the island.}
	\label{S5}
\end{figure*}

Figure \ref{S5} shows thickness dependence. Below a certain thickness the two BM modes switch places in frequency. When $\textrm{BM}_{3}$ has a higher frequency than $\textrm{BM}_{1}$ this suggests magnetostatic surface spin wave (MSSW) behaviour rather than backward volume magnetostatic spin-wave (BVMSSW) behaviour at larger thickness. There are several avoided crossings in this heat-map. Hybridisation is apparent in modes 3 and 4 which was seen for the `chevron' array in the main manuscript in Fig. 4. Another avoided-crossing occurs around 26 nm where a higher order EM (10) intersects $\textrm{BM}_{1}$. Similarly mode 9 shows some mode avoided-crossing at around 36 nm. What these modes all have in common is significant stray-field.

\begin{figure*}
	\centering
		\includegraphics[width=0.6\textwidth]{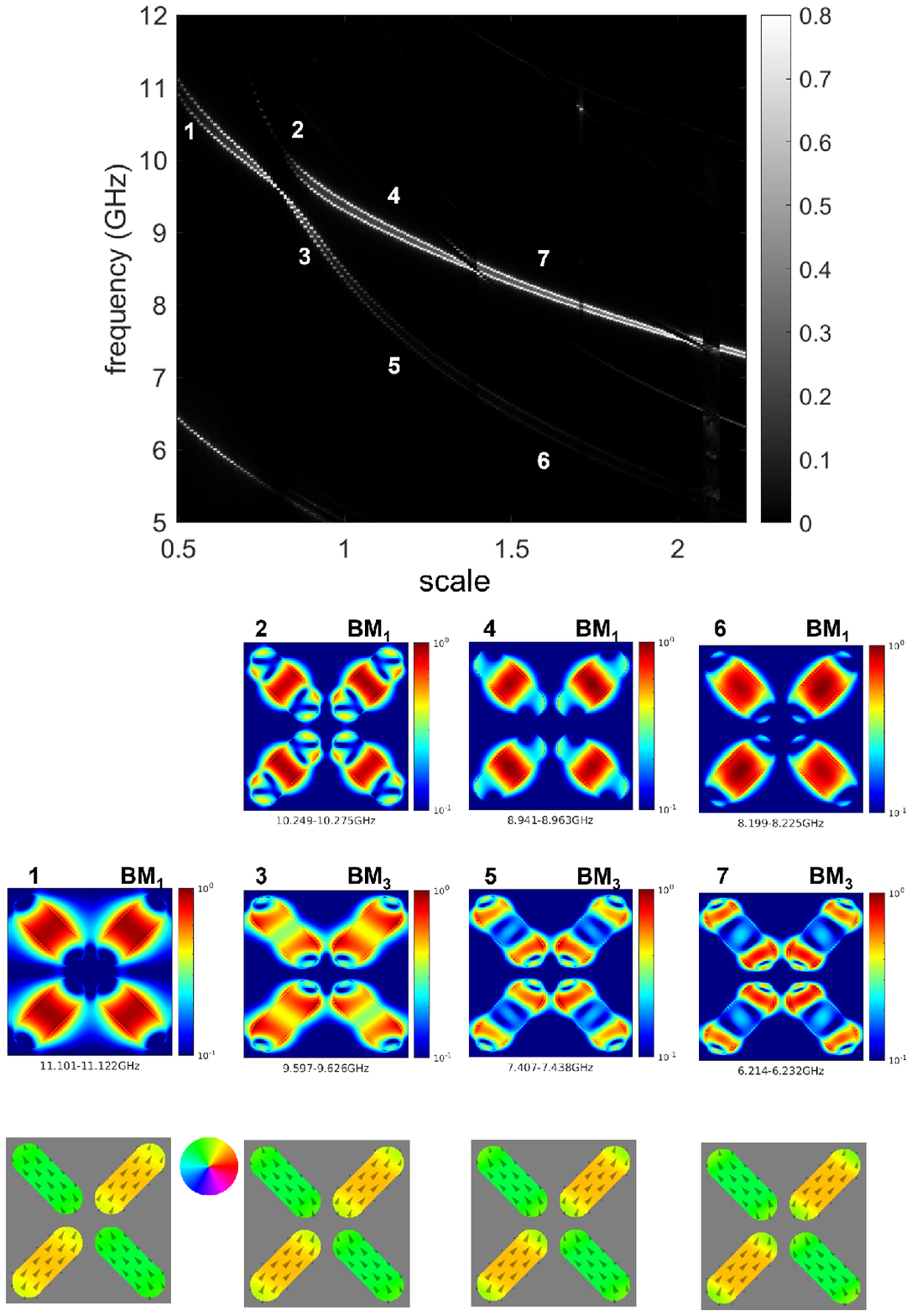}
	\caption{Power maps for Fig. 5d in the main manuscript in which lateral scale is investigated. Numbers correspond to modes in the heat map. Magnetisation texture is shown underneath colour corresponding to the colour wheel. $\textrm{BM}_{1}$ and $\textrm{BM}_{3}$ hybridise around 0.8}
	\label{S6}
\end{figure*}

Figure \ref{S6} shows how the mode behaviour changes with lateral scale. Modes 2 and 3 show signs of hybridisation between $\textrm{BM}_{1}$ and $\textrm{BM}_{3}$. The amount of stray-field emanating from the nanoislands clearly drops as the scale is increased.

\bibliography{references}